\documentclass[a4paper,12pt]{article}

\newcommand{\cA}{\mathcal{A}}
\newcommand{\cB}{\mathcal{B}}
\newcommand{\cC}{\mathcal{C}}
\newcommand{\cD}{\mathcal{D}}
\def\a  {\alpha}
\def\b  {\beta}
\def\g  {\gamma}
\def\d  {\delta}

\def\la {\lambda}
\def\r  {\rho}

\def\G  {{\mit\Gamma}}
\def\D  {{\mit\Delta}}

\def\La {{\mit\Lambda}}

\newcommand{\dfrac}[2]{\displaystyle{\frac{#1}{#2}}}
\newcommand{\dsum}{\displaystyle\sum}
\newcommand{\dprod}{\displaystyle\prod}

\newcommand{\tr}{\mathrm{tr}\,}
\usepackage{epsfig}
\title{M\"obius Symmetry of Discrete Time Soliton Equations}
\author{Jun-ichi YAMAMOTO\footnote{E-mail: yjunichi@phys.metro-u.ac.jp}\\
{\small Department of Physics, Tokyo Metropolitan University} \\ 
{\small Minamiohsawa 1-1, Hachiohji, Tokyo 192-0397 Japan.}
}
\date{}
\begin{document}
\sloppy
\maketitle
\begin{abstract}
We have proposed, in our previous papers\cite{NSSY,SSYY}, a method to characterize integrable discrete soliton equations. In this paper we generalize the method further and obtain a $q$-difference Toda equation, from which we can derive various $q$-difference soliton equations by reductions. 
\end{abstract}
%////////////////////////////////////////////////////////////////////%
%
%  Section 1 : Introduction
%
%////////////////////////////////////////////////////////////////////%
\section{Introduction}

It has been known that, for a given soliton equation, there exists a discrete analogue of the evolution equation, which preserves integrability. Once we find a discrete integrable equation, we can derive infinitely many integrable differential equations by taking continuous limits of variables in many different ways. We are interested in characterizing such integrable discrete systems.

In our previous papers \cite{NSSY,SSYY} we studied various integrable discrete systems which have certain symmetry and satisfy periodic boundary conditions. In this approach the nature of a quadratic equation, imposed by the boundary condition, plays a crucial role. The map which generates a time evolution turns out to be twofold, corresponding to two solutions of the quadratic equation. We showed that the discriminant of the quadratic equation becomes a perfect square when the system is integrable. Owing to this fact the map is free from a square root, hence is rational. We will call this type of map a non square root map (NSRM) in what follows. The discrete versions of Lotka-Volterra equation, KdV equation, Toda lattice, KP equation and Painlev\'e equations \cite{HTI,OHT,HT,JM,Kup,Su,KNY,Masuda} are such examples. The purpose of this paper is to generalize this scheme of characterizing discrete integrable systems.

To clarify the point of our argument let us show first a typical example. The 3-point discrete time Toda lattice is given by the set of coupled equations:
\begin{equation}
\left\{
\begin{array}{ccc}
\bar{x}_{k}\bar{u}_k&=&u_kx_{k-1}\\
\bar{x}_k+\bar{u}_{k+1}&=&u_k+x_k
\label{eqn:dToda}
\end{array}
\right.,\qquad k=1,2,3\ .
\end{equation}
Here $\bar x_k$ means the variable $x_k$ at the time $t+1$, {\it i.e.}, $\bar x_k(t)=x_k(t+1)$. We impose the periodic conditions $x_{k+3}=x_k$ and $u_{k+3}=u_k$. Hence there are 6 dependent variables.

Before solving eq.(\ref{eqn:dToda}) directly it is more instructive to solve
\begin{equation}
\left\{
\begin{array}{ccc}
\bar{x}_{k}\bar{u}_k&=&p_k\\
\bar{x}_k+\bar{u}_{k+1}&=&q_k
\label{eqn:dToda2}
\end{array}
\right.
\end{equation}
for arbitrary functions $p_k,\ q_k$ of $(x,u)$. We can also write them as
\begin{equation}
\left\{
\begin{array}{ccc}
\bar{x}_{k}&=&\displaystyle{{q_k\bar x_{k+1}-p_{k+1}\over\bar x_{k+1}}}\\
\bar{u}_k&=&\displaystyle{{q_{k-1}\bar u_{k-1}-p_{k-1}\over\bar u_{k-1}}}.
\label{eqn:dToda3}
\end{array}
\right.
\end{equation}
Note that they are special cases of M\"obius transformation. Hence by repeating this transformation, say for the first equation of eq.(\ref{eqn:dToda3}), three times, $\bar x_k$ is given in terms of $(\bar x_{k+3},\ \bar u_{k-3})$, which are nothing but $(\bar x_k,\ \bar u_k)$ themselves owing to the periodic boundary conditions. In this way we obtain an equation
\begin{equation}
\bar{x}_{k}=\displaystyle{{A\bar x_{k}+B\over \G\bar x_{k}+\D}}
\label{eqn:dToda4}
\end{equation}
or, equivalently,
\begin{equation}
\G\bar x_k^2-(A-\D)\bar x_k-B=0,
\label{quadratic eq}
\end{equation}
where
\begin{eqnarray*}
A&=& q_1q_2q_3-q_kp_k-q_{k-1}p_{k-1},
\\
B&=& -q_kq_{k+1}p_{k+1}+p_{k+1}p_{k-1},\\
\G&=& q_{k+1}q_{k-1}-p_k,\\
\D&=& -q_{k+1}p_{k+1},
\label{ABCD}
\end{eqnarray*}
in the case of three point Toda lattice.

The quadratic equation eq.(\ref{quadratic eq}) has two solutions
\begin{equation}
\bar x_k={A-\D\pm\sqrt{Dis}\over 2\G},
\end{equation}
unless its discriminant
\begin{equation}
Dis=(A-\D)^2+4B\G
\end{equation}
vanishes. The general form of solutions contain square root of polynomial functions. Under this circumstance a sequence of the map will yield very complicated orbits in general. But the integrable systems are not such cases, {\it i.e.}, the discriminant happens to be a perfect square of some polynomial. In fact if we substitute the right hand sides of eq.(\ref{eqn:dToda}) into $p_k$'s and $q_k$'s of the expression eq.(\ref{ABCD}), the discriminant is considerably simplified and is given by
\begin{equation}
Dis=(x_1x_2x_3-u_1u_2u_3)^2.
\end{equation}
This is a typical non square root map (NSRM), which we are going to study 
(see below for more precise definition).

This remarkable feature could be understood from other point of view as well. Namely by an inspection of our original equations eq.(\ref{eqn:dToda}) we can see easily that
\begin{equation}
\bar x_k=u_k,\qquad \bar u_k=x_{k-1}
\label{x=u}
\end{equation}
is one of sets of solution. This implies that another set of solutions of the quadratic equation eq.(\ref{quadratic eq}) must not have a square root either. We would like to emphasize that the existence of the solution of such simple form eq.(\ref{x=u}) owes to the symmetry of the equation eq.(\ref{eqn:dToda}).

We will study, in the following section, $N$-point maps under which the square roots are removed when periodic boundary conditions are imposed. In section 3 the discrete time Toda equation is studied in detail based on the preceding argument. In particular we will show that this scheme enables one to generalize the Toda equation to a $q$-difference form, so that various $q$-difference soliton equations are derived by reductions, including the $qP_{IV}$ equation by Kajiwara, Noumi and Yamada\cite{KNY}. In section 4, we will discuss time evolution of solutions. The last section is devoted to summarize arguments in this paper.

%////////////////////////////////////////////////////////////////////%
%
%  Section 2 : N-point Map
%
%////////////////////////////////////////////////////////////////////%
\section{$N$-point Map with Periodic Boundary Conditions}

We consider $N$-point systems in this section and study conditions under which a map becomes a non square root map (NSRM).

% 2.1 : Mobius //////////////////////////////////////////////////////%
\subsection{M\"obius transformation}

Generalizing eq.(\ref{eqn:dToda3}), let us consider the following sequence of M\"obius transformations,
\begin{equation}
\bar{x}_{k+h(n-1)}=\frac{\a_{n}\bar{x}_{k+hn}+\b_{n}}{\g_{n}\bar{x}_{k+hn}+\d_{n}},\qquad
n=1,2,\cdots, N,\quad h=\pm 1,\quad  \a_{n}\d_{n}-\b_{n}\g_{n}\neq 0\ 
\label{1-Step-Moebius}
\end{equation}
and M\"obius matrices 
\begin{equation}
M_n=\left(
\begin{array}{cc}
\a_n & \b_n \\
\g_n & \d_n
\end{array}
\right),\quad
\det M_n\neq 0,\qquad n=1,2,\cdots, N.
\end{equation}
After $N$ times of the transformation we obtain the equation
\begin{equation}
\bar x_k={A\bar x_{k+hN}+B\over \G\bar x_{k+hN}+\D}\ ,
\label{N step map}
\end{equation}
while the corresponding M\"obius map is given by
\begin{equation}
S_N:=\left(\begin{array}{cc}A&B\\ \G&\D\end{array}\right)=M_1M_2\cdots M_{N}.
\label{S=MM...M}
\end{equation}

We now impose the periodic boundary conditions $\bar x_k=\bar x_{k+hN}$. Then the formula eq.(\ref{N step map}) is equal to 
the quadratic equation
\begin{equation}
\G\bar x_k^2-(A-\D)\bar x_k-B=0.
\label{quadra eq}
\end{equation}
The solutions of this equation are given by
\begin{equation}
\bar{x}_k=\frac{\tr S_N - 2\D \pm \sqrt{Dis}}{2\G}
\label{sol:quad}
\end{equation}
whereas the discriminant $Dis$ is 
\begin{equation}
Dis=(A-\D)^2+4B\G=(\tr S_N)^2-4\det S_N.
\label{dis}
\end{equation}

% 2.2 : NSRM  //////////////////////////////////////////////////////%
\subsection{Non square root maps}

In this subsection we study maps such that the discriminant eq.(\ref{dis}) becomes a perfect square of some polynomial 
function. As we will prove shortly the following 6 patterns of the map $M_n$ satisfy the requirement:
$$
\left(\begin{array}{cc} 0 & \b_n \\ \b_n & 0 \end{array}\right),\quad
\left(\begin{array}{cc} \a_n & 0 \\ 0 & \d_n \end{array}\right),\quad
\left(\begin{array}{cc} \a_n & \b_n \\ 0 & \d_n \end{array}\right),
$$
\begin{equation}
\left(\begin{array}{cc} \a_n & 0 \\ \g_n & \d_n \end{array}\right),\quad
\left(\begin{array}{cc} \g_n+\b_n & -\b_n \\ \g_n & 0 \end{array}\right),\quad
\left(\begin{array}{cc} 0 & \b_n \\ -\g_n & \b_n+\g_n \end{array}\right).
\label{nonSRS-pat}
\end{equation}

\noindent
Proof:

Since the trace and the determinant are invariant under the transposition ot the matrices, we do not consider here the 
fourth and sixth patterns. For the first three patterns we find 
\begin{equation}
S_N=
\b \left(\begin{array}{cc} 0 & 1 \\ 1 & 0\end{array}\right)^N,\quad 
\left(\begin{array}{cc} \a & 0 \\ 0 & \d \end{array}\right),\quad 
\left(\begin{array}{cc} \a & F \\ 0 & \d \end{array}\right)
\end{equation}
where
$$
\a=\a_1\cdots \a_{N},\quad \b=\b_1\cdots \b_{N},\quad \d=\d_1\cdots \d_{N},
$$
and $F$ is a certain function of $\a_n,\b_n,\d_n$. From this expression we easily find
$$
Dis =
(2\b)^2\delta_{N,odd},\qquad
(\a-\d)^2,\qquad
(\a-\d)^2
$$
corresponding to the first, second and the third patterns.

For the 5th pattern we first observe 
\begin{equation}
\det S_N=\det M_1 \cdots \det M_{N}=\b\g.
\label{det of 5th}
\end{equation}
In order to calculate $\tr S_N$, we split $M_n$ into two parts
\begin{equation}
M_n=\left(\begin{array}{cc} \b_n & -\b_n \\ 0 & 0 \end{array}\right)
+\left(\begin{array}{cc} \g_n & 0 \\ \g_n & 0 \end{array}\right)
\equiv B_n+C_n.
\label{split:5th}
\end{equation}
Using the property
$$
B_mC_n=0,\qquad m,n\in 1,2,\cdots N,
$$
we can see that $S_N$ has terms of the form $C_1\cdots C_\r B_{\r+1}\cdots B_{N}
$ only. From this fact, the trace of $S_N$ is calculated as
\begin{eqnarray}
\tr S_N &=& \sum_{\r \in N}\tr (C_1\cdots C_\r B_{\r+1}\cdots B_{N}) \\\nonumber
&=& \sum_{\r \in N}\tr (C_2\cdots C_\r B_{\r+1}\cdots B_{N}C_1) \\\nonumber
&=& \tr (B_1\cdots B_{N}) + \tr (C_1\cdots C_{N})\\\nonumber
&=& \b+\g
\label{trace of 5th}
\end{eqnarray}
where $\g=\g_1\cdots \g_{N}$. Similarly we can treat the 6th pattern and obtain the same results as eqs.(\ref{det of 5th}) and (\ref{trace of 5th}) for the determinant and trace, respectively.

Hence we have obtained discriminants for all six patterns as follows:
\begin{center}
\begin{tabular}{lllcl}
1)&$\tr S_N = \delta_{N,even} \b$,&$\det S_N =- \b^2$&$\to$&$Dis=(2\b)^2\delta_{N,odd}$,\\
2)&$\tr S_N = \a+\d$,&$\det S_N = \a\d$&$\to$&$Dis=(\a-\d)^2$,\\
3)& $\tr S_N = \a+\d$,&$\det S_N = \a\d$&$\to$&$Dis=(\a-\d)^2$,\\
4)& $\tr S_N = \a+\d$,&$\det S_N = \a\d$&$\to$&$Dis=(\a-\d)^2$, \\
5)& $\tr S_N = \g+\b$,&$\det S_N = \b\g$&$\to$&$Dis=(\g-\b)^2$, \\
6)& $\tr S_N = \g+\b$,&$\det S_N = \b\g$&$\to$&$Dis=(\g-\b)^2$. \\
\end{tabular}\end{center}

Therefore every discrete equation with the M\"obius matrix pattern of eq.(\ref{nonSRS-pat}) has no square root, although they are not all.

% 2.3 : Rational Maps ////////////////////////////////////////////////%
\subsection{Rational maps}

Solutions of the quadratic equation eq.(\ref{quadra eq}) are given by eq.(\ref{sol:quad}). The purpose of this subsection 
is to write them explicitly in the case of NSRM. First we note that the 2nd and the 3rd pattern as well 
as even $N$ case of the 1st pattern do not form a quadratic equation due to $\G=0$. Hence it is enough to consider the 
odd $N$ case of 1st pattern, 4th, 5th and 6th patterns.

% 2.3.1 : 1st ////////////////////////////////////////////////////////%
\subsubsection{Odd $N$ 1st pattern}
The odd $N$ case is given by the following setup:
\begin{equation}
S_N=
\left(\begin{array}{cc} 0 & \b \\ \b & 0 \end{array}\right),\quad
\tr S_N=0,\quad
\det S_N=-\b^2,\quad
Dis = (2\b)^2.
\end{equation}
Hence the solutions of the quadratic equation are 
\begin{equation}
\bar{x}_k=
\frac{\tr S_N - 2\D \pm \sqrt{Dis}}{2\G}=
\frac{\pm 2\b }{2\b}=\pm 1.
\label{sol:1st-odd}
\end{equation}
% 2.3.2 : 4th ////////////////////////////////////////////////////////%
\subsubsection{4th pattern}
In this pattern we have
\begin{equation}
S_N=
\left(\begin{array}{cc} \a & 0 \\ F & \d \end{array}\right),\quad
\tr S_N=\a+\d,\quad
\det S_N=\a\d,\quad
Dis = (\a-\d)^2.
\end{equation}
To calculate $F$, we split $M_n$ to two parts $A_n$ and $D_n$ as follows:
\begin{equation}
M_n
=\left(\begin{array}{cc} \a_n & 0 \\ \g_n & \d_n \end{array}\right)
=\left(\begin{array}{cc} \a_n & 0 \\ 0 & 0 \end{array}\right)
+\left(\begin{array}{cc} 0 & 0 \\ \g_n & \d_n \end{array}\right)
\equiv A_n+D_n.
\end{equation}
Using the fact $A_mD_n=0$, we see that $S_N$ has terms of the form $D_1\cdots D_{\r-1} A_{\r}\cdots A_{N}$ only, where 
$\r=1,2,\cdots,N+1$ and $D_0=A_{N+1}=0$. 
All possible terms can be written in such forms as
\begin{eqnarray}
A_1A_2\cdots A_{N} &=& 
\left(\begin{array}{cc} \a & 0 \\ 0 & 0 \end{array}\right), \\
D_1D_2\cdots D_{N} &=&
\left(\begin{array}{cc} 0 & 0 \\ \d_1\cdots \d_{N-1}\g_N & \d \end{array}\right), \\
D_1\cdots D_{\r-1}A_{\r}\cdots A_{N} &=& 
\left(\begin{array}{cc} 0 & 0 \\ \d_1\cdots \d_{\r-2}\g_{\r-1}\a_\r\cdots \a_N & 0\end{array}\right) .
\end{eqnarray}
Hence we find the expression of $F$ as
\begin{equation}
F=\sum_{\r=2}^N\d_1\cdots \d_{\r-2}\g_{\r-1}\a_\r\cdots \a_N.
\end{equation}
The solutions of the quadratic equation are then given as follows:
\begin{equation}
\bar{x}_k=
\frac{\tr S_N - 2\D \pm \sqrt{Dis}}{2\G}=
\frac{(\a-\d) \pm (\a-\d)}{2F}=
\left\{
\begin{array}{c}
\dfrac{\a-\d}{F}\\ \\
0
\end{array}
\right. .
\label{sol:4th}
\end{equation}

% 2.3.3 : 5th ////////////////////////////////////////////////////////%
\subsubsection{5th pattern}
We already got $\tr S_N$ and $\det S_N$ in the previous subsection, but did not have calculated the elements of $S_N$. 
To obtain the solution of the quadratic equation, we must find their elements since information about $C$ and $D$ are 
necessary for the solutions.

The splitting idea of eq.(\ref{split:5th}) is convenient to obtain the elements here again. Using this idea
, we need to calculate only the following quantities
\begin{eqnarray}
C_1C_2\cdots C_{N}&=&\g
\left(\begin{array}{cc} 1 & 0 \\ 1 & 0 \end{array}\right),\\
B_1B_2\cdots B_{N}&=&\b
\left(\begin{array}{cc} 1 & -1 \\ 0 & 0 \end{array}\right),\\
C_1\cdots C_{\r-1}B_{\r}\cdots B_{N}
&=&\g_1\cdots \g_{\r-1}\b_\r\cdots \b_N
\left(\begin{array}{cc} 1 & -1\\ 1 & -1 \end{array}\right),
\end{eqnarray}
because $B_mC_n=0$. Hence $\G$ and $\D$ of the elements of $S_N$ are obtained as
\begin{eqnarray}
\G&=&
 \g_1\cdots \g_N+\sum_{\r=2}^{N}\g_1\g_2\cdots \g_{\r-1}\b_{\r}\cdots \b_N,\\
\D&=&
 -\sum_{\r=2}^{N}\g_1\g_2\cdots \g_{\r-1}\b_\r\cdots \b_N.
\end{eqnarray}
Consequently, noting the fact $\G=\g-\D$, we obtain the solutions of the quadratic equation as
\begin{eqnarray}
\bar{x}_k&=&
\frac{\tr S_N - 2\D \pm \sqrt{Dis}}{2\G}=
\frac{(\g+\b)-2\D \pm (\g-\b)}{2(\g-\D)} \nonumber\\
\nonumber\\
&=&
\left\{
\begin{array}{ccc}
1 \\ \\
\dfrac{\b - \D }{\g-\D}&\equiv&\dfrac{\b_{N}}{\g_1}\dfrac{Q_{N}(\b,\g)}{Q_1(\b,\g)}
\end{array}
\right.
\label{sol:5th}
\end{eqnarray}
where $Q_{n}(\b,\g)$ is the homogeneous polynomial 
of $(\b_1,\cdots,\b_N,\g_1,\cdots,\g_N)$, that is
\begin{equation}
Q_n(\b,\g)=Q_n(\b_1,\cdots,\b_N,\g_1,\cdots,\g_N)=\sum_{\r=2}^{N}\g_{n+1}\cdots \g_{n+\r-1}\b_{n+\r}\cdots \b_{n+N-1}.
\label{def:Q-poly}
\end{equation}

% 2.3.4 : 6th ////////////////////////////////////////////////////////%
\subsubsection{6th pattern}
For this pattern too, we already got $\tr S_N$ and $\det S_N$ in the previous subsection, but did not have calculated 
the elements of $S_N$. Using the splitting idea again, the elements of $S_N$ in this pattern are able to be obtained. 
The splitting of $M_n$ in this case is
\begin{equation}
M_n
=\left(\begin{array}{cc} 0 & \b_n \\ -\g_n & \b_n+\g_n \end{array}\right)
=\left(\begin{array}{cc} 0 & \b_n \\ 0 & \b_n \end{array}\right)
+\left(\begin{array}{cc} 0 & 0 \\ -\g_n & \g_n \end{array}\right)
\equiv B_n+C_n.
\end{equation}
Since $C_mB_n=0$, we are enough to calculate the following terms
\begin{eqnarray}
C_1C_2\cdots C_{N}&=&\g
\left(\begin{array}{cc} 0 & 0 \\ -1 & 1 \end{array}\right),\\
B_1B_2\cdots B_{N}&=&\b
\left(\begin{array}{cc} 0 & 1 \\ 0 & 1 \end{array}\right),\\
B_1\cdots B_{\r-1}C_{\r}\cdots C_{N}
&=&\b_1\b_2\cdots \b_{\r-1}\g_\r\cdots \g_N
\left( \begin{array}{cc} -1 & 1\\ -1 & 1 \end{array}\right).
\end{eqnarray}
Hence we obtain $\G$ and $\D$ of the elements of $S_N$ as
\begin{eqnarray}
\G&=&-\g-\sum_{\r=2}^N\b_1\cdots \b_{\r-1}\g_\r\cdots \g_N,\cr
\D&=&\b+\g+\sum_{\r=2}^{N}\b_1\cdots \b_{\r-1}\g_\r\cdots \g_N.
\end{eqnarray}
If we further use $\G=\b-\D$, we get solutions of the quadratic equation as follows:
\begin{eqnarray}
\bar{x}_k&=&
\frac{\tr S_N - 2\D \pm \sqrt{Dis}}{2\G}=
\frac{(\g+\b)-2\D \pm (\b-\g)}{2(\b-\D)} \nonumber\\
\nonumber\\
&=&
\left\{
\begin{array}{ccc}
1\\ \\
\dfrac{\D-\g}{\D-\b}&\equiv&\dfrac{\b_{1}}{\g_{N}}\dfrac{Q_{1}(\g,\b)}{Q_{N}(\g,\b)}.
\end{array}
\right.
\label{sol:6th}
\end{eqnarray}

% 2.3.5 : Q-polynomial //////////////////////////////////////////////%
\subsection{Polynomial $Q$}
We note that polynomial $Q$'s, which appear in the solutions of the 5th and 6th patterns, are given by the minor determinants of the diagonal entries of the following matrix.
\begin{equation}
Q(\b_1,\cdots,\b_N,\g_1,\cdots,\g_N):=\left(
\begin{array}{ccccccc}
	\b_1+\g_1&\b_1&0&0&\cdots&0&\g_1 \\
	\g_2&\b_2+\g_2&\b_2&0&\cdots&0&0 \\
	0&\g_3&\b_3+\g_3&&&&\vdots \\
	\vdots&&&&&&0 \\
	0&0&\cdots&&&&\b_{N-1} \\
	\b_N&0&\cdots&&0&\g_N&\b_N+\g_N
\end{array}
\right) .\label{Q}
\end{equation}
For simplicity, we will introduce the following notations.
$$
\begin{array}{lcl}
Q(\b,\g)&:=&Q(\b_1,\cdots,\b_N,\g_1,\cdots,\g_N),\\
Q(\g,\b)&:=&Q(\g_1,\cdots,\g_N,\b_1,\cdots,\b_N),\\
Q^\vee(\b,\g)&:=&Q(\b_N,\cdots,\b_1,\g_N,\cdots,\g_1),\\
Q^\vee(\g,\b)&:=&Q(\g_N,\cdots,\g_1,\b_N,\cdots,\b_1).
\end{array}
$$
Namely, the matrix $Q$ having entries in ascending order is expressed by $Q$, and in descending order by $Q^\vee$. These matrices are related by
$$
Q^\vee(\b,\g) = JQ(\g,\b)J,\qquad Q(\b,\g) = JQ^\vee(\g,\b)J
$$
where
\begin{equation}
J:=\left(
\begin{array}{cccccccccc}
0 & \cdots & \cdots & 0 & 1 \\
\vdots &   &        & 1 & 0 \\
\vdots &   & \cdots &   & \vdots \\
0      & 1 &        &   & \vdots \\
1      & 0 & \cdots & \cdots & 0 
\end{array}
\right).
\end{equation}

We set symbols of them of $(n,n)$-entry as $Q_n$, {\it i.e.} 
$$
\begin{array}{cl}
Q_n(\b,\g)&=
\dsum_{\r=1}^N\dprod_{\mu=1}^{\r-1}\g_{n+\mu}\dprod_{\nu=\r}^{N-1}\b_{n+\nu},\\
Q_n(\g,\b)&=
\dsum_{\r=1}^N\dprod_{\mu=1}^{\r-1}\b_{n+\mu}\dprod_{\nu=\r}^{N-1}\g_{n+\nu},\\
Q_n^\vee(\b,\g)&=
\dsum_{\r=1}^N\dprod_{\mu=1}^{\r-1}\g_{n+N-\mu}\dprod_{\nu=\r}^{N-1}\b_{n+N-\nu},\\
Q_n^\vee(\g,\b)&=
\dsum_{\r=1}^N\dprod_{\mu=1}^{\r-1}\b_{n+N-\mu}\dprod_{\nu=\r}^{N-1}\g_{n+N-\nu},
\end{array}
$$
where $\b_{n\pm N}=\b_n,\ \g_{n\pm N}=\g_n$. 

These polynomials satisfy the following useful relations.
\begin{enumerate}
	\item $Q_{N-n+1}(\b,\g)=Q_n^\vee(\g,\b),\quad Q_n(\b,\g)=Q_{N-n+1}^\vee(\g,\b)$,\label{Q=Q^v} \vspace{2mm}
	\item $(\b_n+\g_n)Q_n(\b,\g)=\b_{n-1}Q_{n-1}(\b,\g)+\g_{n+1}Q_{n+1}(\b,\g)$\label{(b+c)Q=bQ+cQ}.
\end{enumerate}
The relation $\ref{Q=Q^v}$ is obvious. The relation $\ref{(b+c)Q=bQ+cQ}$ is non-trivial and will be checked by means of more fundamental relation :
$$
\b_nQ_n(\b,\g)=\b-\g+\g_{n+1}Q_{n+1}(\b,\g)
$$
where $\b=\b_1\cdots\b_N,\;\g=\g_1\cdots\g_N$.

We can show that $Q_n(\beta,\gamma)$ coincides with the cofactor of the $(n,n)$ element of the product $RL$ of the following matrices
$$
L=\left(
\begin{array}{ccccccccc}
	1&0&0&\cdots&\b_N\\
	\b_1&1&&0\\
	0&\b_2&1&\cdots&\\
	&&&\vdots&\\
	0&0&\cdots&\b_{N-1}&1
\end{array}
\right),\quad
R=\left(
\begin{array}{ccccccccc}
	\g_1&1&0&\cdots&0\\
	0&\gamma_2&1&&\\
	&&\vdots&&\\
	0&&&&1\\
	1&0&\cdots&0&\g_N
\end{array}
\right)
$$
%which were used in \cite{HTI} to represent the dToda equations.
which could represent the dToda equations \cite{HTI}.

% 2.3.6 : Modification ///////////////////////////////////////////////%
\subsection{Modification}

The discriminant $Dis$ is invariant under arbitrary similarity transformations, since it is given by the trace and 
determinant of $S_N$. This fact enables us to modify our map starting from the six patterns we considered above.

Namely under the transformation of $M_n$, 
\begin{equation}
M_n \to M_n^{mod}\equiv U_n^{-1}M_nU_{n+1},
\label{transf:mod}
\end{equation}
the matrix eq.(\ref{S=MM...M}) is changed into
\begin{equation}
S_N^{mod}=M_1^{mod}M_2^{mod}\cdots M_{N}^{mod}
=U_1^{-1}M_1M_2\cdots M_{N}U_{N+1}.
\end{equation}
If we take into account the periodic boundary condition, we must have $U_{N+1}=U_1$. Therefore the trace and the 
determinant of $S_N$, hence the discriminant $Dis$ as well, remain constant:
\begin{equation}
\tr S_N^{mod}=\tr S_N,\quad \det S_N^{mod}=\det S_N,\quad Dis^{mod}=Dis,
\end{equation}
under the transformation eq.(\ref{transf:mod}) for arbitrary matrices $U_n$'s. This implies that the modified map 
$M_n^{mod}$ is NSRM if $M_n$ is so. 

Now denoting by $\bar{x}^{mod}$ the modified variable of $\bar{x}$, we will write the M\"obius map associated with $M_n$ 
as $M_n(x_k)$. In this way, the modified M\"obius map is possible to be written as
\begin{equation}
\bar{x}^{mod}_k=M_1^{mod}(\bar{x}_{k+1}^{mod})=U_1^{-1}M_1U_1(\bar{x}_{k+1}^{mod}).
\end{equation}
Here if we define $\bar{x}_{k+n}^{mod}=U_{n+1}^{-1}(\bar{x}_{k+n})$, then
\begin{equation}
\begin{array}{cccl}
&\bar{x}_k^{mod}&=&U_1^{-1}M_1U_2(\bar{x}_{k+1}^{mod})\\
\Rightarrow
&U_1^{-1}(\bar{x}_k)&=&U_1^{-1}M_1U_2U_2^{-1}(\bar{x}_{k+1})\\ 
\Rightarrow
&\bar{x}_k&=&M_1(\bar{x}_{k+1}).
\end{array}
\end{equation}
As a result, the following two modifications are equivalent
\begin{equation}
\begin{array}{clclcl}
1) & M_n & \to & M_n^{mod} &=& U_{n}^{-1}M_nU_{n+1},\\ 
2) & \bar{x}_{k+n} & \to & \bar{x}_{k+n}^{mod}&=&U_{n+1}^{-1}(\bar{x}_{k+n}).
\end{array}
\end{equation}
This correspondence is preserved in the solutions of the quadratic equation, 
so we can create many equations from the basic six patterns in this way.

%////////////////////////////////////////////////////////////////////%
%
%  Section 3 : dToda
%
%////////////////////////////////////////////////////////////////////%
\section{Discrete Time Toda Equation} \label{sec:GDTE}

We study, in this section, the $N$-point discrete time Toda equation (dToda) in detail from the view point of the previous section.

% 3.1 : 2-type of solution ///////////////////////////////////////////%
\subsection{Two types of solution}

Let us start with writing the $N$-point dToda equation.
\begin{equation}
\left\{
\begin{array}{ccc}
\bar{x}_{k}\bar{u}_k&=&u_kx_{k-1}\\
\bar{x}_k+\bar{u}_{k+1}&=&u_k+x_k
\label{eqn:N-point dToda}
\end{array}
\right.,\qquad k=1,2,3,\cdots, N.
\end{equation}
In order to apply the argument in the previous section to this system we transform eq.(\ref{eqn:N-point dToda}) into the form of M\"obius transformation,
\begin{eqnarray}
\left\{
		\begin{array}{lcl}
		\bar{x}_k&=&\dfrac{(x_k+u_k)\bar{x}_{k+1}-x_ku_{k+1}}{\bar{x}_{k+1}}\\
		\bar{u}_k&=&\dfrac{(x_{k-1}+u_{k-1})\bar{u}_{k-1}-x_{k-2}u_{k-1}}{\bar{u}_{k-1}}
		\end{array}
	\label{MM:Toda 1}
	\right.
\end{eqnarray}
corresponding to $(\bar x_{k+1}, \bar u_{k-1})\rightarrow (\bar x_k,  \bar u_{k})$. 

We notice that this is a modified 5th pattern M\"obius transformation which is specified by the following data:
\begin{equation}
(\b_n, \g_n)=\left\{
\begin{array}{c}
(x_{k+n-1},u_{k+n-1})\\
(u_{k-n},x_{k-n})
\end{array}\right.,
\qquad
U_n=\left(
\begin{array}{cc}
1 & 0 \\
0 & \g_n
\end{array}
\right).
\label{SetUp:Toda}
\end{equation}
Substituting these data into eq.(\ref{sol:5th}) we obtain solutions immediately. Before writing them down, however, it is convenient to simplify notations. Namely the matrix $Q$, which appears in the solutions, is $Q(x_k,x_{k+1},\cdots,x_{k-2},x_{k-1},u_k,u_{k+1},\cdots,u_{k-2},u_{k-1})$. %The minors satisfy
Let the matrix $Q$ be 
$$
\begin{array}{lll}
Q(x_{k},u_{k})&:=&Q(x_k,\cdots,x_{k+N-1},u_k,\cdots,u_{k+N-1}),\\
Q^\vee(x_{k},u_{k})&:=&Q(x_{k+N-1},\cdots,x_k,u_{k+N-1},\cdots,u_k),\\
\end{array}
$$
and its minor determinants be
\begin{equation}
\begin{array}{lll}
Q_n(x_k,u_k)&:=&Q_n(x_k,\cdots,x_{k+N-1},u_k,\cdots,u_{k+N-1}),\\
Q_n^\vee(x_k,u_k)&:=&Q_n(x_{k+N-1},\cdots,x_k,u_{k+N-1},\cdots,u_k).
\end{array}
\end{equation}
These minor determinants satisfy
\begin{enumerate}
	\item $Q_{N-n+1}(x_k,u_k)=Q^\vee_n(u_k,x_k),\quad Q_n(x_k,u_k)=Q^\vee_{N-n+1}(u_k,x_k)$,\vspace{2mm}
	\item $(x_{k+n}+u_{k+n})Q_n(x_k,u_k)=x_{k+n-1}Q_{n-1}(x_k,u_k)+u_{k+n+1}Q_{n+1}(x_k,u_k),$\vspace{2mm}
	\item $Q_n(x_{k+N},u_{k+N})=Q_{n+1}(x_{k+N-1},u_{k+N-1})=\cdots=Q_{n+N-1}(x_{k+1},u_{k+1})=Q_{n+N}(x_k,u_k)$
\end{enumerate}
where $x_{k\pm N}=x_k,\;u_{k\pm N}=u_k$. 
Therefore the solutions turn out to be
\begin{equation}
\left\{
\begin{array}{lc}
	\left\{\begin{array}{l}
	\bar{x}_k=x_{k-1}\dfrac{Q_N(x_k,u_k)}{Q_1(x_k,u_k)}\vspace{3mm}\\
	\bar{u}_k=u_{k}\dfrac{Q_1(x_k,u_k)}{Q_N(x_k,u_k)}
	\end{array}\right.
	&\textrm{A-type},
%\label{A-type:Toda}
\vspace{5mm}\\
	\left\{\begin{array}{l}
	\bar{x}_k=u_k\\
	\bar{u}_k=x_{k-1}
	\end{array}\right.
	&\textrm{B-type}.
\label{B-type:Toda}
\end{array}
\right.
\end{equation}

Similarly the time reversal map is obtained by writing eq.(\ref{eqn:N-point dToda}) as
\begin{eqnarray}
\left\{
		\begin{array}{lcl}
		\begin{array}{lcl}
		x_k&=&\dfrac{(\bar{x}_k+\bar{u}_{k+1})x_{k-1}-\bar{x}_k\bar{u}_k}{x_{k-1}}\\
		u_k&=&\dfrac{(\bar{x}_k+\bar{u}_{k+1})u_{k+1}-\bar{x}_{k+1}\bar{u}_{k+1}}{u_{k+1}}
		\end{array}
		\end{array}
	\label{MM:Toda 1 inverse}
	\right.
\end{eqnarray}
corresponding to the map $(x_{k-1}, u_{k+1}) \rightarrow (x_k, u_k)$.

This is again a modified 5th pattern of the M\"obius transformation. The modification matrix and the elements of the map are given by
\begin{equation}
(\b_n, \g_n)=
	\left\{
		\begin{array}{l}
		(\bar{x}_{k+1-n},\bar{u}_{k+2-n})\\
		(\bar{u}_{k+n},\bar{x}_{k-1+n})
		\end{array}
	\right.
%\label{SetUp:Toda inverse}
,\qquad
U_n=\left(
\begin{array}{cc}
1 & 0 \\
0 & \g_n
\end{array}
\right),
\label{mtxU:Toda:5th inverse}
\end{equation}
and the solutions
\begin{equation}
\left\{
	\begin{array}{lc}
	\left\{
		\begin{array}{l}
		x_{k-1}=\bar{x}_k\dfrac{Q_N^\vee(\bar{x}_k,\bar{u}_{k+1})}
								{Q_1^\vee(\bar{x}_k,\bar{u}_{k+1})}
		\vspace{3mm}\\
		u_k=\bar{u}_{k}\dfrac{Q_1^\vee(\bar{x}_k,\bar{u}_{k+1})}
								{Q_{N}^\vee(\bar{x}_k,\bar{u}_{k+1})}
		\end{array}
		\right.
	&\textrm{A-type},
%\label{A-type:Toda inv}
\vspace{5mm}\\
\left\{
\begin{array}{l}
x_k=\bar{u}_{k+1}\vspace{0mm}\\
u_k=\bar{x}_k
\end{array}
\right.
&\textrm{B-type}.
\label{B-type:Toda inv}
\end{array}
\right.
\end{equation}

Instead of eq.(\ref{MM:Toda 1}) and eq.(\ref{MM:Toda 1 inverse}) we could write the dToda equation eq.(\ref{eqn:N-point dToda}) in the form
\begin{eqnarray}
		&&\left\{
		\begin{array}{lcl}
		\bar{x}_k&=&\dfrac{-x_{k-1}u_k}{\bar x_{k-1}-(x_{k-1}+u_{k-1})}\\
		\bar{u}_k&=&\dfrac{x_{k-1}u_{k}}{\bar{u}_{k+1}-(x_k+u_k)}
		\end{array}
	\right. ,
\label{MM:Toda 2}
	\\
	&&\left\{
		\begin{array}{lcl}
		x_k&=&\dfrac{-\bar{x}_{k+1}\bar{u}_{k+1}}{x_{k+1}-(\bar{x}_{k+1}+\bar{u}_{k+1})}\\
		u_k&=&\dfrac{-\bar{x}_k\bar{u}_{k}}{u_{k-1}-(\bar{x}_{k-1}+\bar{u}_{k})}
		\end{array}
	\right. 
\label{MM:Toda 2 inverse}
\end{eqnarray}
corresponding to $(\bar x_{k-1}, \bar u_{k+1})\rightarrow(\bar x_k, \bar u_k)$ and $(x_{k+1}, u_{k-1})\rightarrow(x_k, u_k)$.
These are modified 6th patterns whose elements and the modification matrix are given by
\begin{equation}
(\b_n, \g_n)=\left\{
	\begin{array}{ll}
	\left\{
		\begin{array}{l}
		(x_{k-1-n},u_{k-1-n})\\
		(u_{k+n},x_{k+n})
		\end{array}
	\right.\\
	\left\{
		\begin{array}{l}
		(\bar{x}_{k+1+n},\bar{u}_{k+1+n})\\
		(\bar{u}_{k-n},\bar{x}_{k-1-n})
		\end{array}
	\right.
	\end{array}
	\right.
	,\qquad
%\label{SetUp:Toda 2}
U_n=\left(
\begin{array}{cc}
1/\g_{n-1} & 0 \\
0 & 1
\end{array}
\right)
\label{mtxU:Toda:6th}
\end{equation}
for eq.(\ref{MM:Toda 2}) and  eq.(\ref{MM:Toda 2 inverse}), respectively.
 Applying these data to eq.(\ref{sol:6th}) we obtain the same results of eq.(\ref{B-type:Toda}) and eq.(\ref{B-type:Toda inv}).

% 3.2 : Generalizations ///////////////////////////////////////////%
\subsection{Generalizations}

% 3.2.1 : B-type //////////////////////////////////////////////////%
\subsubsection{Symmetry of B-type solutions}
We have seen that the B-type solution is simply 1 if the map $M_n$ is the form of either 5th or 6th pattern. A modification of the matrix by $U_n$ will change the solutions as well as the equation while the discriminant remains constant. In particular a different B-type solution will correspond to a different set of equations, and hence different A-type solution. If the B-type solution is rational the A-type solution must be also rational satisfying the same equations. 

To see how it works we try to generalize the B-type solution eq.(\ref{B-type:Toda}) to
\begin{equation}
\left\{\begin{array}{l}
\bar x_k=u_{k+i}\\
\bar u_k=x_{k+j}
\end{array}\right.
,\qquad i,j=0,1,2,\cdots, N-1.
\end{equation}
It is not difficult to find their corresponding equations
\begin{equation}
\left\{\begin{array}{l}
\bar x_{k+l}\bar u_k=x_{k+j}u_{k+i+l}\\
\bar x_k+\bar u_{k+m}=x_{k+j+m}+u_{k+i}
\end{array}\right.
,\qquad l,m=0,1,2,\cdots, N-1.\label{gdToda}
\end{equation}
From the construction this equation has also A-type solution which is a rational polynomial of $x$ and $u$. 
And we can set $h=m+l=\pm 1$ without loss of generality.

We would like to emphasize that the B-type solution follows automatically to the symmetry possessed by the equation. In other words, the rational solution of A-type is a consequence of the symmetry of the Toda lattice equation.

% 3.2.2 : q-difference ///////////////////////////////////////////////%
\subsubsection{$q$-difference version}

In the paper \cite{KNY} Kajiwara {\it et al.} have studied affine-Weyl symmetry of the discrete time Painlev\'e IV equation with parameters, which is called $q$P$_{IV}$. We have shown in our previous paper \cite{SSYY} that $q$P$_{IV}$ is one of the equations which are characterized by the square root free maps. 

Now we want to know whether exist similar generalization of the discrete time Toda lattice eq.(\ref{eqn:N-point dToda}) which we have been studying. To this end we consider the following equation with a set of parameters $a_k, b_k,\  k=1,2,\cdots,N$ :
\begin{equation}
\left\{
\begin{array}{ccc}
\dfrac{\bar{x}_{k+l}}{a_{k+l}}\dfrac{\bar{u}_k}{b_k}&=&a_{k+j}x_{k+j}b_{k+i+l}u_{k+i+l}\\
\dfrac{\bar{x}_k}{a_k}+\dfrac{\bar{u}_{k+m}}{b_{k+m}}&=&a_{k+j+m}x_{k+j+m}+b_{k+i}u_{k+i} .
\label{qToda}
\end{array}
\right.
\end{equation}
Let us call this equation a $q$-difference generalized discrete time Toda equation ($q$Toda). Remarkably this equation has the symmetry of modified 5th or 6th pattern of the M\"obius map. So, this is transformed to a quadratic equation of every $\bar{x},\bar{u},x$ or $u$, and has two types of map as solutions to eq.(\ref{qToda}). 

In fact if we introduce 
\begin{eqnarray}
M_n&=&\left(
\begin{array}{cc}
a_{k+s+nh}x_{k+s+nh}+b_{k+i+nh}u_{k+i+nh} & -a_{k+s+nh}x_{k+s+nh} \\
b_{k+i+nh}u_{k+i+nh} & 0
\end{array}
\right),\\
U_n&=&\left(
\begin{array}{cc}
1 & 0 \\
0 & b_{k+i+nh}u_{k+i+nh}
\end{array}
\right),
\end{eqnarray}
we obtain two maps as follows :
\begin{equation}
\frac{\bar{x}_k}{a_k}=
\left\{
\begin{array}{ll}
a_{k+s-h}x_{k+s-h}\dfrac{Q_N^h(a_{k+s}x_{k+s},b_{k+i}u_{k+i})}{Q_1^h(a_{k+s}x_{k+s},b_{k+i}u_{k+i})},& \textrm{A-type} \\ \\
b_{k+i}u_{k+i},& \textrm{B-type}
\end{array}
\right. \label{qGDTodaMap0}
\end{equation}
where $s=j+m,h=l+m$, and $Q_n^h$ is defined by
$$
Q_n^h(x_k,u_k):=\left\{
\begin{array}{lcl}
Q_n(x_k,u_k)& \textrm{if} & h=1 \\
Q_n^\vee(x_k,u_k)& \textrm{if} & h=-1. \\
\end{array}
\right.
$$

% 3.2.3 : Higher Dim. ///////////////////////////////////////////////%
\subsubsection{Higher dimension}

We can further generalize $q$Toda to higher dimensions by increasing the number of suffices: 
$$
(x_k, u_k)\to (x_{k_1,k_2,\cdots, k_M},\ u_{k_1,k_2,\cdots, k_M}).
$$
The B-type map should have the form
\begin{equation}
\left\{
\begin{array}{ccc}
\displaystyle{{\bar x_{k_1,k_2,\cdots, k_M}\over a_{k_1,k_2,\cdots, k_M}}}&=&b_{k_1+i_1,k_2+i_2,\cdots, k_M+i_M}u_{k_1+i_1,k_2+i_2,\cdots, k_M+i_M}\\\\
\displaystyle{{\bar u_{k_1,k_2,\cdots, k_M}\over b_{k_1,k_2,\cdots, k_M}}} &=&a_{k_1+j_1,k_2+j_2,\cdots, k_M+j_M}x_{k_1+j_1,k_2+j_2,\cdots, k_M+j_M}.
\end{array}
\right.
\end{equation}

In the case of $M=2$, for instance, 
\begin{equation}
\left\{
\begin{array}{ccc}
\displaystyle{{\bar x_{k,l}\over a_{k,l}}}&=&b_{k+1,l+1}u_{k+1,l+1}\\\\
\displaystyle{{\bar u_{k,l}\over b_{k,l}}}&=&a_{k+1,l-1}x_{k+1,l-1}
\end{array}
\right.
\end{equation}
satisfies
\begin{equation}
\left\{
\begin{array}{ccc}
\displaystyle{{\bar x_{k-1,l}\bar u_{k-1,l}\over a_{k-1,l}b_{k-1,l}}}
&=&
a_{k,l-1}b_{k,l+1}x_{k,l-1}u_{k,l+1},\\\\
\displaystyle{{\bar x_{k,l}\over a_{k,l}}}+{\bar u_{k-1,l+1}\over b_{k-1,l+1}}
&=&
a_{k,l}x_{k,l}+b_{k+1,l+1}u_{k+1,l+1}.\\
\end{array}
\right.
\end{equation}
This is the $q$KP equation which we derived in our previous paper.

% 3.3 : Reduction ///////////////////////////////////////////////////%
\subsection{Reductions}

It has been well known that the discrete time Lotka-Volterra equation (dLV) can be obtained from the discrete time Toda equation by a Miura transformation \cite{HT}. We like to show, in this section, that the same dLV-type as well as the discrete time KdV (dKdV)-type equations can be derived from the discrete time Toda lattice eq.(\ref{qToda}) by simple reductions. The basic idea of our reduction is to deal with one of two equations in $q$Toda as a constraint to eliminate half of the variables.

% 3.3.1 : dLV-type //////////////////////////////////////////////////%
\subsubsection{Reduction to the dLV-type}

The way of reduction to a dLV-type equation is to impose an additional constraint to the second equation of $q$Toda as follows.
\begin{equation}
\left\{
\begin{array}{ccccc}
\dfrac{\bar{x}_{k+l}}{a_{k+l}}\dfrac{\bar{u}_k}{b_k}&=&a_{k+j}x_{k+j}b_{k+i+l}u_{k+i+l}\vspace{2mm}\\
\dfrac{\bar{x}_k}{a_k}+\dfrac{\bar{u}_{k+m}}{b_{k+m}}&=&a_{k+j+m}x_{k+j+m}+b_{k+i}u_{k+i} &=& -r_k 
\end{array}
\right.
\end{equation}
where $r_k$'s are constants. 

This constraint enables us to eliminate variables, say $u$ and $\bar u$, from the first equation. Namely the following substitution
\begin{equation}
\left\{
\begin{array}{ccc}
\dfrac{\bar{u}_{k+m}}{b_{k+m}} &\to& -\dfrac{\bar{x}_k}{a_k}-r_k \\
b_{k+i}u_{k+i} &\to& -a_{k+j+m}x_{k+j+m}-r_k
\end{array}
\right.
\label{substitution}
\end{equation}
yields the dLV-type equation
\begin{equation}
\dfrac{\bar{x}_{k+l}}{a_{k+l}}\left(r_{k-m}+\dfrac{\bar{x}_{k-m}}{a_{k-m}}\right)=a_{k+j}x_{k+j}\left(r_{k+l}+a_{k+j+m+l}x_{k+j+m+l}\right).
\end{equation}

When $r_k=a_k=b_k=1,\ k=1,2,\cdots, N$, this is the dLV equation. Various types of $q$LV equations which were discussed in our previous paper, are included in this equation. 

It is straightforward to obtain two maps of this equation. We find
\begin{equation}
\frac{\bar{x}_k}{a_k}=
\left\{
\begin{array}{ll}
a_{k+s-h}x_{k+s-h}\dfrac{Q_N^h(a_{k+s}x_{k+s},-r_k-a_{k+j+m}x_{k+j+m})}{Q_1^h(a_{k+s}x_{k+s},-r_k-a_{k+j+m}x_{k+j+m})},& \textrm{A-type} \\ \\
-r_k-a_{k+j+m}x_{k+j+m},& \textrm{B-type}
\end{array}
\right. \label{Map:LV-type}
\end{equation}
which could be also obtained from the map of qToda eq.(\ref{qToda}) by the substitution eq.(\ref{substitution}). The $q$P$_{\mathrm{IV}}$ of Kajiwara, Noumi, Yamada \cite{KNY} is included as one of the A-type map.

% 3.3.2 : dKdV-type /////////////////////////////////////////////////%
\subsubsection{Reduction to the dKdV-type}

Let us try an alternative case of constraint. Namely we impose a constraint to the first equation of eq.(\ref{qToda}). 
\begin{equation}
\left\{
\begin{array}{ccccc}
\dfrac{\bar{x}_{k+l}}{a_{k+l}}\dfrac{\bar{u}_k}{b_k}&=&a_{k+j}x_{k+j}b_{k+i+l}u_{k+i+l} &=& r_k ,\vspace{2mm}\\
\dfrac{\bar{x}_k}{a_k}+\dfrac{\bar{u}_{k+m}}{b_{k+m}}&=&a_{k+j+m}x_{k+j+m}+b_{k+i}u_{k+i} .
\end{array}
\right.
\end{equation}
This enables us to eliminate $u$ and $\bar u$ from the second equation according to the rule
\begin{equation}
\left\{
\begin{array}{ccc}
\dfrac{\bar{u}_k}{b_k} &\to& \dfrac{a_{k+l}r_k}{\bar{x}_{k+l}} \vspace{2mm}\\
b_{k+i+l}u_{k+i+l} &\to& \dfrac{r_k}{a_{k+j}x_{k+j}}
\end{array}
\right.
\label{qKdV reduction rule}
\end{equation}
from which we obtain
\begin{equation}
\dfrac{\bar{x}_k}{a_k}+\dfrac{a_{k+l+m}r_{k+m}}{\bar{x}_{k+l+m}}=a_{k+j+m}x_{k+j+m}+\dfrac{r_{k-l}}{a_{k+j-l}x_{k+j-l}}.
\label{qdKdV}
\end{equation}
This equation containes the discrete time KdV equation as a special case. Hence we call eq.(\ref{qdKdV}) the $q$KdV equation. The corresponding solutions are given again by simply replacing $u$ and $\bar u$ in the solutions of $q$Toda according to eq.(\ref{qKdV reduction rule}).
\begin{equation}
\frac{\bar{x}_k}{a_k}=
\left\{
\begin{array}{ll}
a_{k+s-h}x_{k+s-h}\dfrac{Q_N^h(a_{k+s}x_{k+s},r_{k-l}/a_{k+j-l}x_{k+j-l})}{Q_1^h(a_{k+s}x_{k+s},r_{k-l}/a_{k+j-l}x_{k+j-l})},&\textrm{A-type} \\ \\
r_{k-l}/a_{k+j-l}x_{k+j-l}. &\textrm{B-type}
\end{array}
\right. \label{qGDTodaMap0}
\end{equation}

% 3.4 : Reflection maps /////////////////////////////////////////////%
\subsection{Reflection maps}
We define some symbols of the two types of map and their inverse maps of dToda as follows, 
$$
\begin{array}{lclclclclcl}
\cA&:&
	\left(\begin{array}{c} x_k \\ u_k \end{array}\right)&\longmapsto&
	\left(\begin{array}{c}
	x_{k-1}\dfrac{Q_N(x_k,u_k)}{Q_1(x_k,u_k)}\vspace{2mm}\\
	u_{k}\dfrac{Q_1(x_k,u_k)}{Q_N(x_k,u_k)}
	\end{array}\right),
&\quad&
\cB&:&
	\left(\begin{array}{c} x_k \\ u_k \end{array}\right)&\longmapsto&
	\left(\begin{array}{c} u_k \\ x_{k-1}\end{array}\right),
\vspace{3mm}\\
\cA^{-1}&:&
	\left(\begin{array}{c} x_{k-1} \\ u_k \end{array}\right)&\longmapsto&
	\left(\begin{array}{l}
x_k\dfrac{Q_N^\vee(x_k,u_{k+1})}{Q_1^\vee(x_k,u_{k+1})}
\vspace{2mm}\\
u_k\dfrac{Q_1^\vee(x_k,u_{k+1})}{Q_N^\vee(x_k,u_{k+1})}
\end{array}\right),
&\quad&
\cB^{-1}&:&
	\left(\begin{array}{c} x_k \\ u_k \end{array}\right)&\longmapsto&
	\left(\begin{array}{l} u_{k+1} \\ x_k \end{array}\right).
\end{array}
$$
These maps satisfy the following relations.
\begin{equation}
\begin{array}{cl}
(\mathrm{I})&\cA\cA^{-1}=\cA^{-1}\cA=1,\ \cB\cB^{-1}=\cB^{-1}\cB=1,\\
(\mathrm{II})&\cA\cB^2=\cB^2\cA,\\
(\mathrm{III})&\cA\cB\cA=\cB^3,
\end{array}
\label{AB}
\end{equation}
or
\begin{equation}
\begin{array}{cl}
(\mathrm{I'})&\cB\cB^{-1}=\cB^{-1}\cB=1,\\
(\mathrm{II'})&\cC\cB=\cB\cD,\\
(\mathrm{III'})&\cC^2=\cD^2=1,
\end{array}
\label{CD}
\end{equation}
where $\cC=\cA^{-1}\cB=\cB^{-1}\cA$ and $\cD=\cA\cB^{-1}=\cB\cA^{-1}$.
The relations eq.(\ref{CD}) are derived from eq.(\ref{AB}) by some calculations, automatically.

The reflection maps $\cC,\;\cD$ change the inverse A-type map to the inverse B-type one and the A-type one to the B-type one, respectively. Namely,
$$
\begin{array}{clcl}
\cC\;: & \textrm{inverse A-type} &\longmapsto& \textrm{inverse B-type},\\
       & \textrm{inverse B-type} &\longmapsto& \textrm{inverse A-type},\\
\cD\;: & \textrm{A-type} &\longmapsto& \textrm{B-type},\\
       & \textrm{B-type} &\longmapsto& \textrm{A-type}.
\end{array}
$$
Hence, these reflection maps are permutations of solutions of the quadratic equation eq.(\ref{quadra eq}) in the case of dToda, {\it i.e.} Galois tranformations of eq.(\ref{quadra eq}). 
Furthermore, the generalized version of dToda eq.(\ref{gdToda}) satisfy these relations eq.(\ref{AB}) or eq.(\ref{CD}), but $q$Toda does not. 
These reflection maps play an important role in affine-Wyel symmetries and $q$-difference Painlev\'e equations as studied by Kajiwara, {\it et al}. \cite{KNY, Masuda}.

%////////////////////////////////////////////////////////////////////%
%
%  Section 4 : Discrete time evolution
%
%////////////////////////////////////////////////////////////////////%
\section{Discrete Time Evolution of the $q$Toda}

A sequence of B-type map does not generate an orbit but causes exchange of the variables $x$ and $u$, whereas an A-type map generates an orbit. We will study a discrete time evolution of A-type map in this section. To make the time dependence explicit we write the variables $\bar x$ and $\bar u$ as $x^{t+1}$ and $u^{t+1}$ in the following.

% 4.1 : A-type maps //////////////////////////////////////////////////%
\subsection{Time evolution of the A-type maps}
A repeated use of eq.(\ref{B-type:Toda}) enables one to express $x^{t+1}_{k}$ and $u^{t+1}_{k}$ in terms of $x^{0}_{k}$'s and $u^{0}_{k}$'s as
\begin{equation}
\left\{
\begin{array}{l}\displaystyle{x^{t+1}_{k}=x_{k-1}^{t}{Q_{k-1}^{t}\over Q_k^{t}}=\cdots =x^0_{k-t-1}\prod_{s=0}^t{Q_{k-s-1}^{t-s}\over Q_{k-s}^{t-s}}},\\
\displaystyle{u^{t+1}_{k}=u_{k}^{t}{Q_{k}^{t}\over Q_{k-1}^{t}}=\cdots =u^0_k\prod_{s=0}^t{Q_{k}^{t-s}\over Q_{k-1}^{t-s}}}.
\end{array}\right.
\label{x^t+1, u^t+1}
\end{equation}
Here we denote by $Q^t_k$ the following polynomial of $x^t$'s and $u^t$'s,
\begin{equation}
Q_k^{t}:=\sum_{j=1}^Nu^{t}_{k+1}u^{t}_{k+2}\cdots u^{t}_{k+j-1}x^{t}_{k+j}\cdots x^{t}_{k+N-1}.
\label{Q_k^t}
\end{equation}

If we substitute eq.(\ref{x^t+1, u^t+1}) into the right hand side of eq.(\ref{Q_k^t}) we will find a considerable cancellation of factors in the numerator and the denominator. The cancellation might be associated with a reduction of algebraic entropy \cite{HV, LRGOT}. As a consequence we obtain a recurrence equation for $Q^t$:
\begin{equation}
Q_k^{t+1}=\sum_{j=1}^Nu^{0}_{k+1}u^{0}_{k+2}\cdots u^{0}_{k+j-1}x^{0}_{k+j-t-1}\cdots x^{0}_{k+N-t-2}
\prod_{s=0}^t{Q^{t-s}_{k+j-1}Q^{t-s}_{k+j-s-1}\over
Q^{t-s}_{k}Q^{t-s}_{k+N-s-1}}.
\label{iteration}
\end{equation}

The A-type solutions are given if $Q^t$ is found by solving eq.(\ref{iteration}) iteratively. The generalization of this result to the case of $q$Toda is straightforward. This is, however, a cumbersome task. And so, we will look orbits of A-type map described by numerical computation.

% 4.2 : Numerical observation ////////////////////////////////////////%
\subsection{Numerical observation of ultra-discrete version of the A-type maps}
It is not difficult to see behaviour of orbits of the A-type map eq.(\ref{x^t+1, u^t+1}) as long as we generate them numerically by using computers. They are constrained on curves since there are sufficient number of conserved quantities. If parameters $a_k$'s and $b_k$'s are introduced the orbits tend to expand and closed curves turn to spirals. These are behaviours which we could expect naively from our previous experiences. Instead of studying them in detail we would like to present here behaviour of ultra-discrete version of solutions.

It has been known that if a discrete integrable equation has its ultra-discrete version the integrability is preserved \cite{TTMS}. There exists a systematic method to ultra-discretize a discrete integrable equation if it contains only product and summation of variables but not subtractions. Namely we introduce a new variable $X$ and replace $x$ by $x=e^{X/\epsilon}$. The ultra-discretization procedure \cite{TTMS} is defined by
$$
\lim_{\epsilon\rightarrow +0}\left(e^{X/\epsilon}\cdot e^{Y/\epsilon}\right)\rightarrow X+Y,\qquad 
\lim_{\epsilon\rightarrow +0}\left(e^{X/\epsilon}+ e^{Y/\epsilon}\right)\rightarrow {\rm max}(X,Y).
$$

To be specific let us consider the case of dKdV-type equation. The $q$-difference version has been given by eq.(\ref{qdKdV}). If we further specify to 3-point case the equations are
\begin{equation}
\dfrac{\bar{x}_k}{a_k}+\dfrac{a_{k+1}}{\bar{x}_{k+1}}=a_{k+1+j}x_{k+1+j}+\dfrac{1}{a_{k+j}x_{k+j}},
\end{equation}
$$
\qquad k=1,2,3, \quad j=0,1,2,\quad a_{k+3}=a_k,\quad x_{k+3}=x_k.
$$
The corresponding A-type map can be also found readily from the one of $q$-Toda equation by the reduction eq.(\ref{qKdV reduction rule}).
\begin{equation}
\bar{x}_{k-j}=
a_{k-j}a_{k+2}x_{k+2}\dfrac{1+a_{k+2}a_{k+1}x_{k+2}x_{k+1}+a_ka_{k+1}^2a_{k+2}x_kx_{k+1}^2x_{k+2}}{1+a_ka_{k+2}x_kx_{k+2}+a_ka_{k+1}a_{k+2}^2x_kx_{k+1}x_{k+2}^2},\label{3-qKdV-A}
\end{equation}
$$
\qquad k=1,2,3, \quad j=0,1,2,\quad a_{k+3}=a_k,\quad x_{k+3}=x_k.
$$
The ultra-discrete version of eq.(\ref{3-qKdV-A}) turns to
\begin{equation}
\begin{array}{cl}
\bar{X}_{k-j}
=&A_{k-j}+A_{k+2}+X_{k+2}\\
&+\max(0,\ A_{k+2}+A_{k+1}+X_{k+2}+X_{k+1},\ 
A_{k}+2A_{k+1}+A_{k+2}+X_{k}+2X_{k+1}+X_{k+2})\\
&-\max(0,\ A_{k}+A_{k+2}+X_{k}+X_{k+2},\ 
A_{k}+A_{k+1}+2A_{k+2}+X_{k}+X_{k+1}+2X_{k+2}).
\end{array}
\end{equation}

%/////// Figure 1, 2 //////////%
\begin{figure}[tbp]
\hspace{.02\linewidth}
	\begin{minipage}[t]{.45\linewidth}
	\begin{center}
		\caption{$A_1=A_2=A_3=0$}
		\epsfysize=7cm
		\centerline{\epsfbox{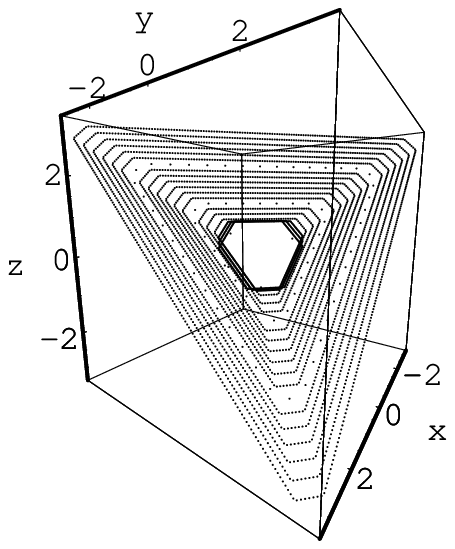}}
		\label{fig:UDKdV00}	
	\end{center}
	\end{minipage}
\hspace{.04\linewidth}
	\begin{minipage}[t]{.45\linewidth}
	\begin{center}
		\caption{$j=0$, $A_1+A_2+A_3=0$}
		\epsfysize=7cm
		\centerline{\epsfbox{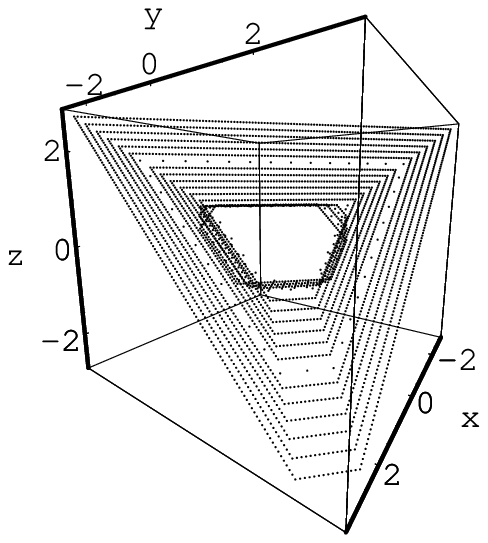}}
		\label{fig:UDKdV01}
	\end{center}
	\end{minipage}
\end{figure}
%/////// Figure 3, 4 //////////%
\begin{figure}[tbp]
\hspace{.02\linewidth}
	\begin{minipage}[t]{.45\linewidth}
	\begin{center}
		\caption{$j=1$, $A_1+A_2+A_3=0$}
		\epsfysize=7cm
		\centerline{\epsfbox{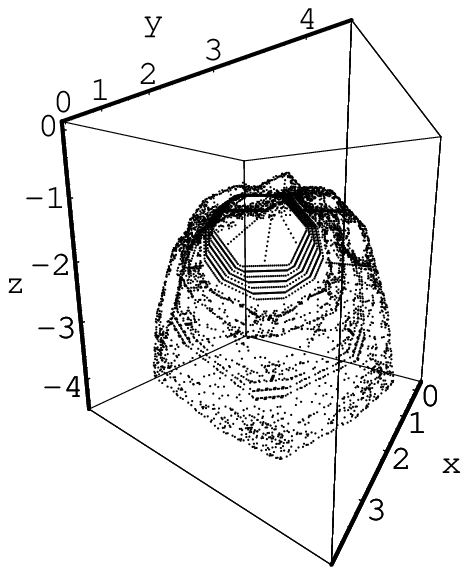}}
		\label{fig:UDKdV02}	
	\end{center}
	\end{minipage}
\hspace{.04\linewidth}
	\begin{minipage}[t]{.45\linewidth}
	\begin{center}
		\caption{$j=2$, $A_1+A_2+A_3=0$}
		\epsfysize=7cm
		\centerline{\epsfbox{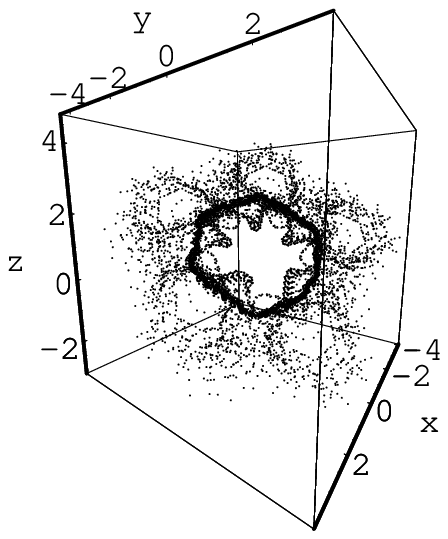}}
		\label{fig:UDKdV03}
	\end{center}
	\end{minipage}
\end{figure}
%///////////////////////////////%
Here we set $a_k=e^{A_k/\epsilon},\ k=1,2,3$. Let $X_k$'s and $A_k$'s be rational numbers and impose a condition $A_1+A_2+A_3=0$ so that the map returns to initial values after some steps. Under these circumstances we find maps such as presented in Figures \ref{fig:UDKdV00} to \ref{fig:UDKdV03}.

Figure 1 shows the case of $A_k=0,\ k=1,2,3$ for all j=0,1,2. Similarly Fig. 2 corresponds to the case of $A_1+A_2+A_3=0$ and $j=0$. The straight lines in these figures are known to correspond to integrable orbits \cite{HIRTGO}. Figures 3 and 4 are the cases of $j=1$ and $j=2$ respectively. They do not seem to describe integrable orbits. Therefore the orbits behave differently for different value of $j$. This must be contrasted with the case of dKdV in which orbits form smooth closed curves under the condition $a_1a_2a_3=1$ for all $j=0,1,2$. 

It will be worth while to point out that the dKdV map eq.(\ref{3-qKdV-A}) in the case of $j=0$ is invariant under the following B\"acklund transformation
$$
\begin{array}{ll}
\pi(a_k)=a_{k+1}, &
\pi(x_k)=x_{k+1}, \vspace{2mm}\\
s_i(a_j)=a_j(a_{i+1}a_{i-1})^{\la_{ij}}, &
s_i(x_j)=x_j\left(\dfrac{a_{i+1}a_{i-1}+x_{i+1}x_{i-1}}{1+a_{i+1}a_{i-1}x_{i+1}x_{i-1}}\right)^{\g_{ij}},
\end{array}
$$
where
\[
\La:=(\la_{ij})=\left(\begin{array}{rrr}2&-1&-1\\-1&2&-1\\-1&-1&2\end{array}\right),\qquad
\G:=(\g_{ij})=\left(\begin{array}{rrr}0&1&-1\\-1&0&1\\1&-1&0\end{array}\right),
\]
\[
s_k^2=1,\qquad
\pi^3=1,\qquad
s_ks_{k+1}s_k=s_{k+1}s_ks_{k+1},\qquad
\pi s_k=s_{k+1}\pi
\]
and $i,j,k=1,2,3$ and $a_{k+3}=a_k,\ x_{k+3}=x_k$. This is the same affine Weyl symmetry shared by $q$P$_{IV}$ \cite{KNY}. But $N$($\ge$4)-point dKdV maps of $j=0$ does not have this symmetry even though it exhibits an integrable orbit.

We have also studied 3-point ultra-discrete LV-type maps in the same manner. The results are similar to those of KdV-type maps and found one integrable map corresponding to the $q$P$_{IV}$.

%////////////////////////////////////////////////////////////////////%
%
%  Section 5 : Summary
%
%////////////////////////////////////////////////////////////////////%
\section{Summary}
The main purpose of this paper was to clarify the method characterizing discrete integrable systems, which we introduced in our porevious works. We have shown that, in the case of integrable systems, two adjacent variables are related  by a M\"obius transformation of special forms. We found at least 6 patterns of such forms. 

This observation enabled us to obtain quite large class of new discrete integrable systems. The generalized discrete Toda equations (dToda) of eq.(\ref{gdToda}) is one of such examples. By certain reductions of the dToda equation dKdV and dLV were obtained. The method admits introduction of arbitrary parameters into the equations and further generalization to $q$Toda systems. 

If we solve the equations in the form in which each variable is given as a function of variables in the previous time, there exist always two types of solutions which we called A-type and B-type. The B-type map does not create orbit but exchanges variables, while a sequence of the A-type map generates an orbit. 

The map of A-type is written as a rational polynomial of variables whose numerator and denominator are expressed by minor determinants of the matrix $Q$ in eq.(\ref{Q}). It is interesting to notice that this matrix $Q$ is the same one known in the field of 
%tropical versions of soliton cellular automata associated with crystal \cite{Ki}, 
affine Weyl group and $q$-Painlev\'e equations \cite{KNY,Masuda}.

We have not succeeded analyzing full behaviour of a sequence of A-type maps. It is one of the most interesting problems to find behaviour of the map when parameters $a_k$'s and $b_k$'s are introduced. Some numerical results using ultra-discrete method have shown that the introduction of the parameters preserve integrability in some cases but not always. This result seems to show that the non square root map condition is not sufficient but necessary for a discrete map being integrable. It is desirable to clarify this point further.

%////////////////////////////////////////////////////////////////////%
%
%  Bibliography
%
%////////////////////////////////////////////////////////////////////%


\begin{thebibliography}{}
\bibitem{NSSY} 
Y.~Narita, S.~Saito, N.~Saitoh and K.~Yoshida : J. Phys. Soc. Jpn. {\bf 70} (2001) 1246.

\bibitem{SSYY}
S.~Saito, N.~Saitoh, J.~Yamamoto and K.~Yoshida : J. Phys. Soc. Jpn. {\bf 70} (2001) 3517.

%\bibitem{Hi} 
%R.~Hirota : Prog. Theor. Phys, {\bf 64} (1975) 288.
%\bibitem{Hirota} 
%R.~Hirota : J.Phys. Soc. Jpn. {\bf 43} (1977) 1424.
%\bibitem{Hirota2}
%R.~Hirota : J. Phys. Soc. Jpn. {\bf 50} (1981) 3785.
\bibitem{HTI}
R.~Hirota, S.~Tsujimoto and T.~Imai : ``{\it Difference Scheme of Soliton Equations}", in {\it Future Directions of Nonlinear Dynamics in Physical and Biological Systems}, ed. by P.L.Christiansen at al., (Plenum Press, New York, 1993) p.7. 
\bibitem{OHT}
Y.~Ohta, R.~Hirota and S.~Tsujimoto : J.Phys. Soc. Jpn. {\bf 62} (1993) 1872.
\bibitem{HT} 
R.~Hirota, and S.~Tsujimoto : J.Phys.Soc.Jpn. {\bf 64} (1995) 3125-3127
\bibitem{JM}
M.~Jimbo and T.~Miwa : J. Phys. Soc. Jpn. {\bf 51} (1982) 4116, 4125. 
\bibitem{Kup}
B.~Kupershmidt : {\it KP or mKP} Mathematical Surveys and Monographs Vol {\bf 78} (American Mathematical Society, 2000) and references therein.
\bibitem{Su} 
Y.~Suris : see, for instance, arXiv:solv-int/9902003.

\bibitem{KNY} 
K.~Kajiwara, M.~Noumi and Y.~Yamada :
%`` {\it A Study on the fourth $q$-Painlev\'e Equation }", 
arXiv:nlin.SI/0012063,

%\bibitem{KNY2} 
Kenji Kajiwara, Masatoshi Noumi and Yasuhiko Yamada :
%``{\it Discrete dynamical systems with $W(A^{(1)}_{m-1} \times A^{(1)}_{n-1})$ symmetry }", 
arXiv:nlin.SI/0106029,

%\bibitem{KNY3} 
Kenji Kajiwara, Masatoshi Noumi and Yasuhiko Yamada :
%``{\it $q$-Painlev\'e systems arising from $q$-KP hierarchy}", 
arXiv:nlin.SI/0112045.

\bibitem{Masuda}
T.~Masuda :
%``{\it On the Rational Solutions of $q$-Painlev\'e V Equation}", 
arXiv:nlin.SI/0107050,

%\bibitem{Masuda2}
T.~Masuda : 
%``{\it On a class of algebraic solutions to Painlev\'e VI equation, its determinant formula and coalescence cascade}", 
arXiv:nlin.SI/0202044.

\bibitem{HV}
J.~Hietarinta and C.Viallet : Phys. Rev. Lett. {\bf 81} (1998) 325, 

\bibitem{LRGOT}
S.~Lafortune, A.Ramani, B.Grammaticos, Y.Ohta and K.M.Tamizhmani : arXiv:nlin.SI/0104020

%\bibitem{TS} D. Takahashi and J. Satsuma,
%``\textit{A soliton cellular automaton}",
%J. Phys. Soc. Jpn. {\bf 59} (1990) 3514--3519.
\bibitem{TTMS}
T. Tokihiro, D. Takahashi, J. Matsukidaira and J. Satsuma :
%``\textit{From Soliton Equations to Integrable Cellular Automata through a Limiting Procedure}", 
Phys. Rev. Lett. {\bf 76} (1996) 3247-3250

\bibitem{HIRTGO}
R. Hirota, M. Iwao, A. Ramani, D. Takahashi, B. Grammaticos and Y. Ohta :
%``\textit{From Integrability to Chaos in a Lotka-Volterra Cellular-Automaton}", 
 Phys. Lett. A {\bf 236} (1997) 39--44 

%\bibitem{Ki} A.N.~Kirillov :
%\textit{Introduction to tropical combinatorics}, 
%in ``Physics and Combinatorics 2000''
%(Eds. A.~N.~Kirillov and N.~Liskova)
%Proceedings of the Nagoya 2000 International Workshop,
%pp. 82--150, World Scientific, 2001.
\end{thebibliography}
\end{document}